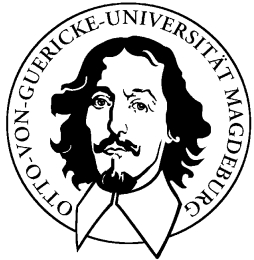

## Institute for Theoretical Physics

## Number partitioning as random energy model

Heiko Bauke†, Silvio Franz‡ and Stephan Mertens†‡
†Institut für Theoretische Physik, Otto-von-Guericke Universität,
PF 4120, 39016 Magdeburg, Germany and
‡The Abdus Salam International Centre for Theoretical Physics,
Condensed Matter Section, Strada Costiera 11, 34014 Trieste, Italy

‡stephan.mertens@physik.uni-magdeburg.de.

# Number partitioning as random energy model

Heiko Bauke†, Silvio Franz‡ and Stephan Mertens†‡

†Institut für Theoretische Physik, Otto-von-Guericke Universität, PF 4120, 39016 Magdeburg, Germany and
‡The Abdus Salam International Centre for Theoretical Physics,
Condensed Matter Section, Strada Costiera 11, 34014 Trieste, Italy

Number partitioning is a classical problem from combinatorial optimisation. In physical terms it corresponds to a long range anti-ferromagnetic Ising spin glass. It has been rigorously proven that the low lying energies of number partitioning behave like uncorrelated random variables. We claim that neighbouring energy levels are uncorrelated almost everywhere on the energy axis, and that energetically adjacent configurations are uncorrelated, too. Apparently there is no relation between geometry (configuration) and energy that could be exploited by an optimization algorithm. This "local random energy" picture of number partitioning is corroborated by numerical simulations and heuristic arguments.

## 1. Introduction

In disordered systems the energy levels $E(\sigma)$ are quenched random variables with distribution induced by the random couplings between the dynamical variables $\sigma$. In general the energy levels are correlated, but under certain conditions these correlations can be neglected. A well known example is the $p$-spin generalisation of the Sherrington-Kirkpatrick model [1], where the correlations decrease with increasing $p$ [2, 3]. In the large $p$ limit, the energy levels can be treated as *independent* random variables, the corresponding model is called *random energy model* or REM [2]. In a REM the role of the dynamical variables $\sigma$ is reduced to that of indices in a table of uncorrelated random energy values.

Considered as an optimisation problem, the REM is as hard as it can get. Locating the minimum in a disordered table obviously requires to search the whole table. In the REM that emerges from the $p$-spin model the size of the table grows exponentially with the number $N$ of spins. Heuristic algorithms that yield near optimal solutions by considering only a polynomial number of configurations usually rely on correlations between configurations and energies, but in a REM no such correlations exist. The large $p$ limit of the $p$-spin model is no prominent problem in combinatorial optimisation, but there is a classical optimisation problem that comes very close to a REM: The *number partitioning problem* (NPP), which we will introduce in the next section. The REM-like nature of the NPP has been heuristically motivated and used as an ansatz to derive the statistics of the optimal and suboptimal configurations [4]. The validity of this approach has been questioned [5], although its predictions have been confirmed rigorously [6]. Apparently there is a need for a more profound discussion of the REM nature of the NPP, and this what we will do in this contribution.

We start with a definition of the NPP and a brief discussion of some of its properties. Section 3 describes the REM approach to the NPP and what has been proven rigorously.

In section 4 we discuss the scope of the REM approach to NPP. Numerical simulations corroborate a conjecture which basically says that any finite number of neighbouring energies becomes uncorrelated in the thermodynamic limit. In section 5 we show that energetically adjacent configurations are uncorrelated, too. The NPP has a natural generalisation called the multiprocessor scheduling problem, and our REM conjecture holds for this generalised problem, too, as we will see in section 6.

## 2. The number partitioning problem

The number partitioning problem is a classic problem from combinatorial optimisation: Given $N$ positive integer numbers $\{a_1, a_2, \ldots, a_N\}$, we seek a partition of these numbers into two subsets such that the sum of numbers in one subset is as close as possible to the sum of numbers in the other subset. The partition can be encoded by Ising spin variables $\sigma_i = \pm 1$, where $\sigma_i = 1$ if $a_i$ is put in one subset and $\sigma_i = -1$ if $a_i$ is put in the other subset. The cost function to be minimised then reads

$$E(\sigma) = \frac{1}{\sqrt{N}} \left| \sum_{i=1}^{N} a_i \sigma_i \right|, \qquad (1)$$

where we have deliberately inserted a factor $1/\sqrt{N}$ to simplify the equations in the rest of the paper.

Number partitioning is of considerable importance, both practically and theoretically. Its practical applications range from task scheduling and the minimisation of VLSI circuit size and delay [7, 8] over public key cryptography [9, 10] to choosing up sides in a ball game [11]. Number partitioning is also one of Garey and Johnson's six basic NP-hard problems that lie at the heart of the theory of NP-completeness [12, 13], and in fact it is the only one of these problems that actually deals with numbers. Hence it is often chosen as a base for NP-completeness proofs of other problems involving numbers, like bin packing, multiprocessor scheduling, quadratic programming or knapsack problems [14]. In physical terms, the optimal partitions can be considered as the ground states of an infinite range, anti-ferromagnetic Ising spin system with

‡stephan.mertens@physik.uni-magdeburg.de.

Mattis-like couplings $J_{ij} = -a_i a_j$ defined by the Hamiltonian

$$H(\sigma) = E^2(\sigma) = \frac{1}{N}\sum_{ij} a_i a_j \sigma_i \sigma_j =: -\frac{1}{N}\sum_{ij} J_{ij}\sigma_i\sigma_j. \quad (2)$$

The statistical mechanics of this model has been discussed by several authors [15, 16, 17, 18].

The computational complexity of the NPP depends on the type of input numbers $\{a_1, a_2, \ldots, a_N\}$. Consider the case that the $a_i$ are positive integers bound by a constant $A$. Then $E(\sigma)$ can take on at most $NA$ different values, i.e. the size of the search space is $NA$ instead of $2^N$ and it is straightforward to devise an algorithm that explores this reduced search space in time polynomial in $NA$ [12]. The NPP is NP-hard only for input numbers of size exponentially large in $N$ or, after division by the maximal input number, of sufficiently high precision. Setting $A = 2^{\kappa N}$ for some $\kappa > 0$ and choosing the $a_i$ randomly and independently from $\{1, 2, \ldots, A\}$, the typical properties of a random instance depend on the value of $\kappa$: For $\kappa < 1$ the ground state entropy per spin is positive, for $\kappa > 1$ it is zero. This *phase transition* in NPP has been studied numerically [19], analytically within a statistical mechanics framework [17] and finally it was analysed rigorously [6]. Similar phase transitions have been observed in many other NP-hard problems. Their study forms the base of an emerging interdisciplinary field of research that joints computer scientists, mathematicians and physicists [20, 21].

## 3. REM approach

For the rest of the paper we consider the NPP with $a_i$ being real numbers, chosen independently and uniformly at random from the interval $(0,1]$, i.e. we consider the "infinite precision limit" or $\kappa \gg 1$. We sort the $2^{N-1}$ different values of the energy $E(\sigma)$ in ascending order,

$$0 \le E_1 \le E_2 \le \ldots \le E_{2^{N-1}}. \quad (3)$$

Borgs, Chayes and Pittel (BCP) proved a remarkable theorem on the statistics of the low lying energies [6, Theorem 2.8]: For any fixed $\ell \ge 1$, the $\ell$-tuple

$$(\varepsilon_1(N), \varepsilon_2(N), \ldots, \varepsilon_\ell(N)) := \sqrt{\frac{6}{\pi}} 2^{N-1} (E_1, E_2, \ldots, E_\ell) \quad (4)$$

converges in distribution to $(w_1, w_1 + w_2, \ldots, w_1 + \cdots + w_\ell)$, where $w_i$ are independent, identically distributed random variables with exponential probability-density

$$p(w) = \exp(-w). \quad (5)$$

The BCP-Theorem implies that the scaled energies

$$\varepsilon_k = \lim_{N\to\infty} \varepsilon_k(N) \quad (6)$$

$(k = 1, 2, \ldots, \ell)$ are random variables with densities

$$p_k(\varepsilon_k) = \frac{\varepsilon_k^{k-1}}{\Gamma(k)} e^{-\varepsilon_k} \quad (7)$$

each. Note that (7) had first been derived by a *random energy approach* [4]. For this approach one *assumes* that energies of the NPP are drawn independently from the density of states,

$$g(E) := \frac{1}{2^N}\sum_\sigma \left\langle \delta\left(E - \frac{1}{\sqrt{N}}|\sum_j a_j \sigma_j|\right)\right\rangle \quad (8)$$
$$\simeq \sqrt{\frac{6}{\pi}}\exp\left(-\frac{3E^2}{2}\right)\Theta(E),$$

where $\langle\cdot\rangle$ denotes the average over $\{a_1, a_2, \ldots, a_N\}$ and the right hand side follows from the central limit theorem. Now assume that we draw $M$ independent numbers from the distribution $g(E)$. We sort these numbers into ascending order (3) and ask for the probability $p_k(E, W)\,\mathrm{d}E\,\mathrm{d}W$ to have $E_k \in [E, E+\mathrm{d}E]$ and $E_{k+1} - E_k \in [W, W+\mathrm{d}W]$. With $G(E) = \int_0^E g(E')\,\mathrm{d}E'$ we get

$$p_k(E,W) = \binom{M}{k} k g(E) [G(E)]^{k-1} \cdot$$
$$(M-k)g(E+W)[1-G(E+W)]^{M-k-1}. \quad (9)$$

For large values of $M$, $p_k(E,W)$ is concentrated at small values of $E$ and $W$. Introducing the scaled variables

$$\varepsilon := M g(0) E \quad \text{and} \quad w := M g(0) W \quad (10)$$

we get

$$p_k(E,W)\,\mathrm{d}E\,\mathrm{d}W \simeq \frac{\varepsilon^{(k-1)}}{\Gamma(k)} e^{-\varepsilon - w}\,\mathrm{d}\varepsilon\,\mathrm{d}w \quad (11)$$

for large $M$. Integration over $\varepsilon$ gives (5) and integration over $w$ gives (7). The REM ansatz leads to the BCP-Theorem provided

$$M g(0) = \sqrt{\frac{6}{\pi}} 2^{N-1}. \quad (12)$$

A naive interpretation of this last equation is $M = 2^{N-1}$, meaning that all energy levels of the NPP are statistically independent. There is no need to run elaborate simulations [5] to see that this assumption of *global* independency is invalid: Configurations with high energies are highly correlated. Consider the configuration with maximum energy (all spins equal). Its nearest neighbour on the energy axis is reached by flipping the spin with the smallest weight. The next nearest neighbours can be reached by only flipping spins with small weights. The spacings of these correlated energy levels are polynomial in $1/\sqrt{N}$, but we are considering levels with spacing $\mathcal{O}(2^{-N})$ here. Instead of global independency we only need a fraction $f$ of low energy levels to be statistically independent in order to derive the BCP-Theorem. Then $M = f 2^{N-1}$ and $g(0) \mapsto g(0)/f$ due to the renormalisation of the underlying probability distribution, and $f$ cancels in (12). The fraction $f$ may depend on $N$, the REM derivation is valid as long as $\lim_{N\to\infty} f 2^{N-1} = \infty$. The REM that is hidden in random NPP is *local*: Only neighbouring energy levels are statistically independent.

## 4. Local random energy model

The REM-like nature of the NPP is counter intuitive at first sight and in fact people have disputed it [5]. The basic mechanism behind the local REM is not hard to understand, however. The ground state energy $E_1$ is $\mathcal{O}\left(2^{-N}\right)$, i.e. its leading $N$ bits are zero. For typical samples this can always be achieved by a careful adjustment of $\sigma$, but the precise value of $E_1$ on the scale $\mathcal{O}\left(2^{-N}\right)$ is determined by the less significant bits of the $a_j$, and this value cannot be controlled by $\sigma$. On a scale $\Delta E = \mathcal{O}\left(2^{-N}\right)$, the noise in the $a_j$ "shines through". Obviously the ground state and its neighbours are not special, and a generalised BCP-Theorem should hold everywhere on the energy axis where the level spacing is $\mathcal{O}\left(2^{-N}\right)$. This is definitely the case for all energies $\mathcal{O}(1)$, which brings us to our local REM conjecture for random NPP: Let $0 \leq E_{N,1} \leq E_{N,2} \leq \ldots \leq E_{N,2^{N-1}}$ denote the sorted list of energies of an instance of the NPP. For convenience we define $E_{N,0} = -1$. Let $\alpha \geq 0$ be an arbitrary, fixed real number, and let $r$ such that

$$E_{N,r} < \alpha \leq E_{N,r+1}, \tag{13}$$

i.e. $E_{N,r}$ is the largest energy smaller than $\alpha$. Then our claim is that for any fixed $\ell \geq 1$, the $\ell$-tuple

$$\sqrt{\frac{6}{\pi}}\, e^{-\frac{3\alpha^2}{2}}\, 2^{N-1}\left((E_{N,r+1}, E_{N,r+2}, \ldots, E_{N,r+\ell}) - \alpha\right) \tag{14}$$

converges in distribution to $(w_1, w_1+w_2, \ldots, w_1+\cdots+w_\ell)$, where $w_i$ are i.i.d. random variables, each distributed exponentially (5). Note that this conjecture is equivalent to a *local random energy ansatz*: The finite sized neighbourhood of each $\mathcal{O}(1)$-energy level consists of statistically independent energy values. Corresponding to (6) we define the scaled energies

$$\varepsilon_{r,i}(N) = \sqrt{\frac{6}{\pi}}\, e^{-\frac{3\alpha^2}{2}}\, 2^{N-1}\left(E_{N,r+i} - \alpha\right) \tag{15}$$

for $i = 1, 2, \ldots, \ell$. Our conjecture implies that

$$\varepsilon_{r,i} = \lim_{N\to\infty} \varepsilon_{r,i}(N) \tag{16}$$

are random variables with densities $p_i(\varepsilon_{r,i})$ (7), and this is confirmed by numerical simulations (figure 1).

With a modification of the Horowitz-Sahni algorithm [22], the values $\varepsilon_{r,i}(N)$ can be found in time $\mathcal{O}\left(N 2^{N/2}\right)$ and space $\mathcal{O}\left(2^{N/2}\right)$. This limits the accessible system sizes to $N < 45$, but luckily the REM scenario can be seen for fairly small values of $N$. Note that the local REM seems to be violated for larger values of $\alpha$, see figure 2. This is a finite size effect: The maximum energy $E_{N,2^{N-1}}$ is $\sqrt{N}/2$ on average, and $\alpha$ must be much smaller than this to sample energies with level spacing $\mathcal{O}\left(2^{-N}\right)$. Asymptotically, the level spacing is $\mathcal{O}\left(2^{-N}\right)$ for all energies $o(\sqrt{N})$, hence we could extend our local random energy hypothesis to energies that increase with $N$, like $E \propto \sqrt{N}/\log N$ for example, but due to the computational limitations and strong finite size effects we could not check this extended claim by numerical simulations.

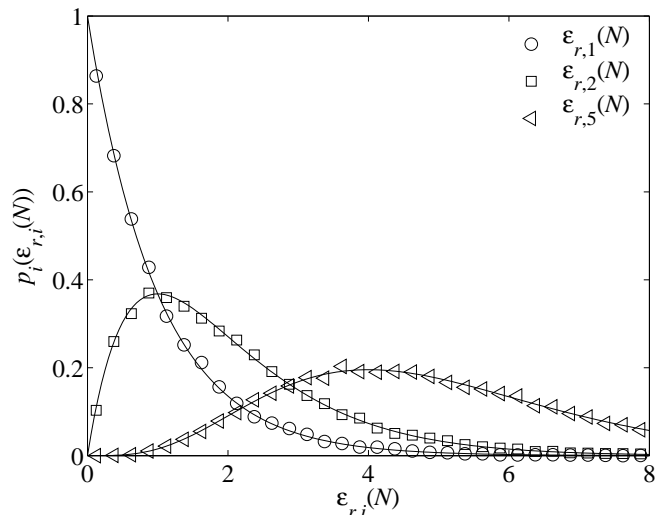

Figure 1: Distribution of the scaled energies $\varepsilon_{r,i}(N)$ for $\alpha = 0.43$. Numerical simulations for $N = 24$ (symbols) and the predictions of the local random energy conjecture, (7) (lines). Each data point represents an average over 25 000 random instances.

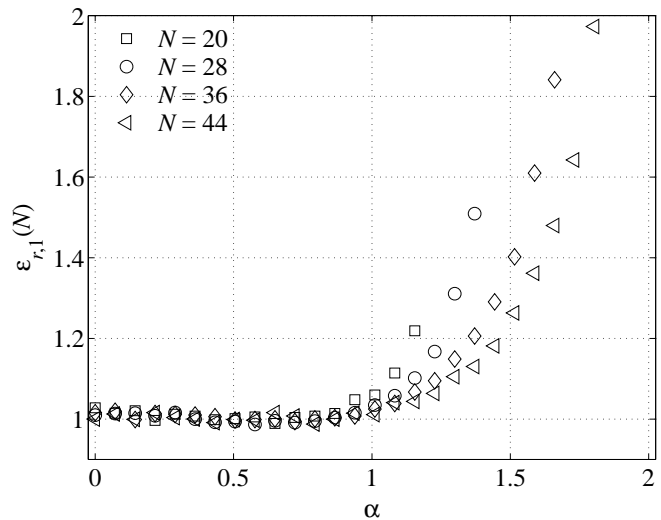

Figure 2: Mean $\overline{\varepsilon_{r,1}(N)}$ as a function of the parameter $\alpha$ for different system sizes $N$. Each data point represents an average over 25 000 random instances.

## 5. Configurations and local REM

So far we have concentrated on the energies, but what about the corresponding configurations? In Derrida's random energy model, configurations and energies are completely disconnected. The same is true in the NPP, at least in a local sense: Energetically adjacent configurations are uncorrelated. The intuitive explanation is this: Imagine that you sit on a configuration $\sigma$ with energy $E(\sigma)$. You know that the nearest level is $\mathcal{O}\left(2^{-N}\right)$ away, but how do you get there in configu-ration space? If you flip a single spin, the energy will change typically by $\mathcal{O}(N^{-1/2})$ and minimally by $\mathcal{O}(N^{-3/2})$. In any case your step size is far too large, and the same holds for each fixed number of subsequent spin flips as $N$ goes to infinity. Reaching your nearest neighbour on the energy axis requires the coordinated flip of $\mathcal{O}(N)$ spins, but the precise set of spins to be flipped depends on the less significant bits in the $a_j$. Again the randomness in the $a_j$ "shines through".

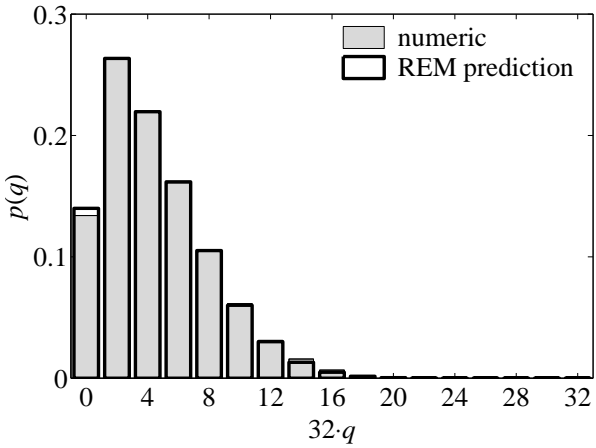

Figure 3: Distribution of the overlap $q$ between the ground state and the the first excited state of the NPP with $N = 32$. The experimental data is based upon 10 000 random samples, the theoretical distribution is given by (18).

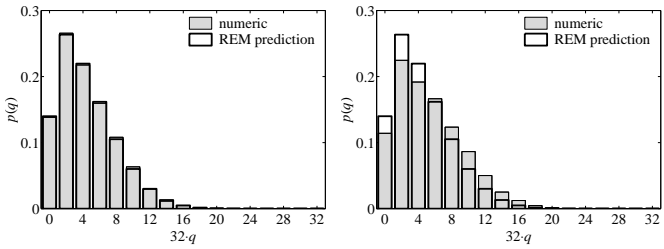

Figure 4: Distribution of the overlap $q$ of two energetically adjacent configurations at energy $\approx \alpha$ with $\alpha = 0.433$ (left) and $\alpha = 1.15$ (right) and $N = 32$ averaged over 10 000 random samples. The theoretical distribution is given by (18).

A simple numerical experiment that corroborates this intuitive picture is to measure the overlap

$$q(\sigma, \sigma') = \frac{1}{N} \left| \sum_{j=1}^{N} \sigma_j \sigma'_j \right| \quad (17)$$

between two configurations $\sigma$ and $\sigma'$ that are neighbours on the energy axis. Performing this measurement for many random instances we get a distribution $p(q)$. If $\sigma$ and $\sigma'$ are uncorrelated, $p(q)$ should be given by the "heads-minus-tails" distribution, the distribution of the absolute difference in the number of heads and tails if a coin is tossed $N$ times,

$$p(q) = \begin{cases} \frac{1}{2^N} \binom{N}{N(1-q)/2} & \text{for } q = 0, \\ \frac{2}{2^N} \binom{N}{N(1-q)/2} & \text{for } q > 0. \end{cases} \quad (18)$$

Figure 3 shows that this is precisely what happens for $\sigma$ being the ground state and $\sigma'$ the first excitation. According to our local REM hypothesis, (18) should not only hold for the ground state and the first excitation but for all configurations with level spacing $\mathcal{O}(2^{-N})$. If we repeat the experiment with configurations that correspond to energies $E_{N,r} < \alpha \leq E_{N,r+1}$, we find the same distribution if $\alpha$ is not too large, see figure 4. Again the deviations for large values of $\alpha$ are finite size effects. Note that the configurations with the two largest energies are separated by a single spin flip only and typical values of these energies are $\sqrt{N}/2$. So for finite systems the REM has to become invalid for large $\alpha$.

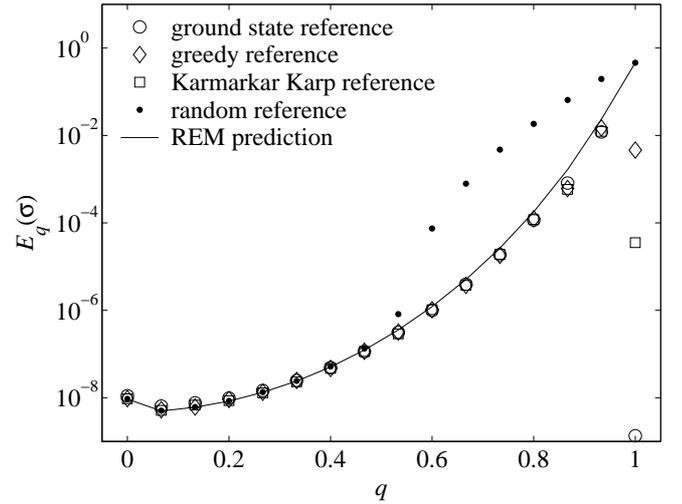

Figure 5: Minimum energy $E_q(\sigma)$ that can be found among configurations that have a fixed overlap $q$ with a reference configuration $\sigma$. Symbols represent averages over 10 000 random samples of size $N = 30$, the solid line is given by (21).

Another experiment [23, 24] that supports the REM nature of the NPP is to calculate the minimum energy $E_q(\sigma)$ that can be found among all configurations that have a given overlap $q$ with a reference configuration $\sigma$,

$$E_q(\sigma) = \min_{\sigma'} \{E(\sigma') \mid q(\sigma, \sigma') = q\}. \quad (19)$$

According to our local REM hypothesis we expect $E_q$ to decrease with the number $M(q)$ of feasible configurations,

$$M(q) = \left(1 - \frac{\delta_{q,0}}{2}\right) \binom{N}{N\frac{1-q}{2}}, \quad (20)$$

or more precisely $E_q(\sigma)$ should be given by the minimum out of $M(q)$ independent random numbers drawn from $g(E)$,

$$E_q = M(q)\sqrt{\frac{6}{\pi}} \int_0^\infty dE\, E \exp\left(-\frac{3E^2}{2}\right) \left[1 - \text{erf}\left(\sqrt{\frac{3}{2}}E\right)\right]^{M(q)-1} \tag{21}$$

$$\simeq \sqrt{\frac{\pi}{6}} \frac{1}{M(q)}.$$

As you can see in figure 5 this behaviour is confirmed by numerical simulations. As reference configuration we have chosen the true ground state, a random configuration or a configuration given by a polynomial time algorithm like the greedy approximation or the Karmarkar-Karp differencing heuristics [25]. The reference energies for these configurations are

$$E_{q=1}(\text{ground state}) = \mathscr{O}\left(2^{-N}\right),$$
$$E_{q=1}(\text{Karmarkar-Karp}) = \mathscr{O}\left(N^{-1/2 - 0.83 \log N}\right),$$
$$E_{q=1}(\text{greedy}) = \mathscr{O}\left(N^{-3/2}\right) \quad \text{and}$$
$$E_{q=1}(\text{random}) = \sqrt{\frac{2}{3\pi}}.$$

All except the random configuration are local minima, hence the very first spin flip increases the energy, but from then on $E_q$ closely follows (21). Apparently a single spin flip out of a local minimum immediately lets the system forget where it came from. This fits nicely with the observation that the NPP behaves like a Trap model under Metropolis dynamics [26]. The deviations observed for random reference configurations are due to the correlations among high energy configurations. A single spin flip can change the energy at most by $2/\sqrt{N}$, but the energy of a random configuration is much larger. The system needs $\mathscr{O}\left(\sqrt{N}\right)$ of steepest descent moves (flip the spin with the largest decrease in energy) before it has forgotten its reference level. According to this considerations the largest value $q$ up to which $E_q(\text{random})$ agrees with the REM prediction (21) should scale like $1 - \mathscr{O}\left(N^{-1/2}\right)$, and this is supported by extended simulations up to $N = 44$.

## 6. Multiprocessor scheduling problem

The NPP has a natural generalisation: Divide a set $\{a_1, a_2, \ldots, a_N\}$ of positive numbers into $q$ subsets such that the sums in all $q$ subsets are as equal as possible. This is known as multiway partitioning or multiprocessor scheduling problem (MSP). The latter name refers to the problem of distributing $N$ tasks with running times $\{a_1, a_2, \ldots, a_N\}$ on $q$ processors of a parallel computer such that the overall running time is minimised.

As for $q = 2$ we assume the weights $a_i$ to be real valued, i.i.d. random numbers from $(0, 1]$. The cost function

$$E(\vec{\sigma}) = \frac{1}{\sqrt{N}} \left| \sum_{i=1}^{N} a_i \vec{\sigma}_i \right| \tag{22}$$

of the MSP is formulated in terms of $(q-1)$-dimensional Potts-vectors $\vec{e}_j$ ($j = 1, \ldots, q$). $\vec{\sigma}_i = \vec{e}_j$ means weight $a_i$ is in subset $j$, see [27] for details. Note that for $q = 2$ (22) reduces to (1).

Generalising the REM ansatz to MSP we assume that the energies of the MSP are drawn independently from the density of states

$$g(E) := q^{-N} \sum_{\vec{\sigma}} \left\langle \delta\left(E - \frac{1}{\sqrt{N}} |\sum_j a_j \vec{\sigma}_j|\right) \right\rangle$$
$$\simeq \left(\frac{3(q-1)}{2}\right)^{\frac{q-1}{2}} \frac{q-1}{\Gamma\left(\frac{q+1}{2}\right)} E^{q-2} \exp\left(-\frac{3(q-1)E^2}{2}\right) \Theta(E), \tag{23}$$

where the last line follows from the central limit theorem. Note that even systems of moderate size are well described by this limiting distribution, see figure 6. Note also the difference between the cases $q = 2$ and $q > 2$: For $q > 2$ the maximum of $g(E)$ is shifted from zero to a value greater than zero, and $g(0) = 0$. This effect has a simple geometric origin [27], but it affects the statistics of the ground states.

It is intuitively clear that the statistics of the ground state is governed by the behaviour of $g(E)$ as $E \to 0$. The rigorous variant of this intuition can be found in any text book on extreme order statistics [28]. It tells us that the distribution of the $i$-th smallest out of $M$ random numbers, drawn independently

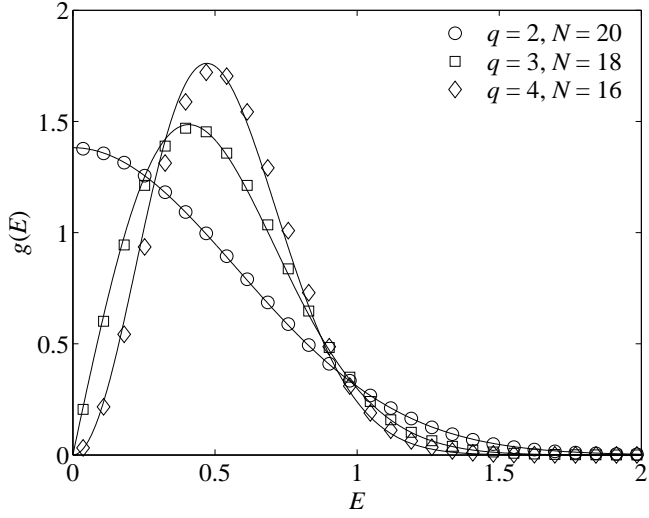

Figure 6: Distribution of the scaled energies $E$ of the MSP with $q = 2, 3, 4$ for finite system sizes. Each histogram was calculated by choosing 100 random instances. Full lines are theoretical distributions (23).

from a distribution $p(x)$ with

$$p(x) = \begin{cases} 0 & x < 0 \\ p_0 x^{\nu-1} + \mathcal{O}(x^\nu) & x \gtrsim 0, \end{cases} \quad (24)$$

becomes a limiting distribution as $M \to \infty$ that depends only on $p_0$ and $\nu$ ($\nu \geq 1$). To be more precise: The distribution of the $i$-th smallest rescaled variable

$$y = \left(M \frac{p_0}{\nu}\right)^{1/\nu} x \quad (25)$$

converges to

$$p_i(y) = \frac{\nu e^{-y^\nu} y^{\nu i - 1}}{\Gamma(i)} \quad (26)$$

as $M \to \infty$. For $i = 1$ this distribution is called Weibull distribution, which for $q = 2$ reduces to the simple exponential distribution. To use this result in our REM ansatz for the MSP we note that

$$\nu = q - 1 \quad \text{and} \quad p_0 = \left(\frac{3(q-1)}{2}\right)^{\frac{q-1}{2}} \frac{q-1}{\Gamma\left(\frac{q+1}{2}\right)}. \quad (27)$$

The number $M$ of energy levels is given by

$$M = \sum_{k=1}^{q} \begin{Bmatrix} N \\ k \end{Bmatrix} \simeq \frac{q^N}{q!}, \quad (28)$$

where $\begin{Bmatrix} N \\ k \end{Bmatrix}$ denotes the number of ways to partition $N$ elements in $k$ non-empty subsets, a quantity usually referred to as Stirling number of the second kind [29].

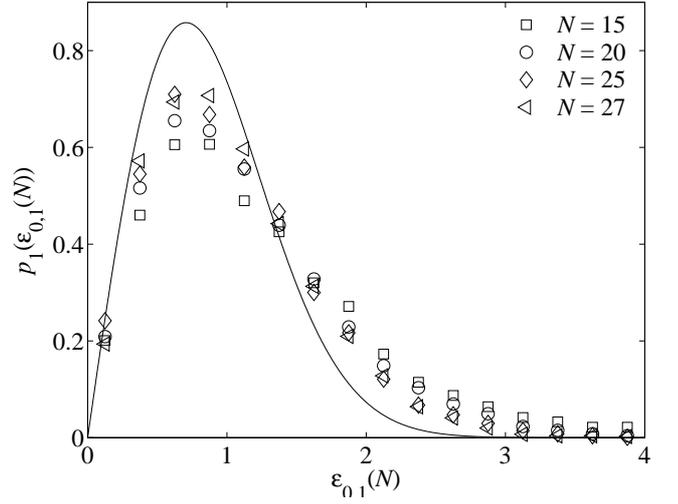

Figure 7: Distribution of the scaled energies $\varepsilon_{0,1}(N)$ of the MSP with $q = 3$. Each histogram was calculated by choosing 5 000 random instances. Solid line is the distribution (30).

Like in the case $q = 2$ we introduce the ordered list $0 \leq E_{N,1} \leq E_{N,2} \leq \ldots$ of energies and their scaled counterpart

$$\varepsilon_{0,i}(N) = \left(\frac{\Sigma_{j=1}^{q} \begin{Bmatrix} N \\ j \end{Bmatrix}}{\Gamma\left(\frac{q+1}{2}\right)}\right)^{\frac{1}{q-1}} \sqrt{\frac{3(q-1)}{2}} E_{N,i}, \quad (29)$$

and (26) yields

$$p_i(\varepsilon_{0,i}) = \frac{(q-1)\varepsilon_{0,i}^{(q-1)i-1}}{\Gamma(i)} \exp\left(-\varepsilon_{0,i}^{q-1}\right) \quad (30)$$

as the asymptotic distribution of these scaled energies, provided the REM-assumption of independent energy levels is true. Figure 7 shows the distribution of $\varepsilon_{0,1}(N)$ for different values of $N$ and $q = 3$ as measured by exact numerical solution of small systems. The deviations from (30) are obvious, but the tendency in the numerical data to approach (30) as $N$ gets larger is also clearly visible. The strong finite size effects for $q > 2$ compared to $q = 2$ are related to the different behaviour of $g(0)$ in both cases. This can be seen if one considers the REM hypothesis at energies $\alpha > 0$ which are more populated than those at $\alpha = 0$.

Again we define $r$ for given $\alpha > 0$ by (13), i.e. $E_{N,r}$ is the largest energy smaller than $\alpha$. Then, for any fixed $\ell \geq 1$, the sequence

$$\varepsilon_{r,i}(N) = Mg(\alpha)(E_{N,r+i} - \alpha) \quad i = 1, 2, \ldots, \ell \quad (31)$$

of subsequent energies or rather their limit

$$\varepsilon_{r,i} = \lim_{N \to \infty} \varepsilon_{r,i}(N) \quad (32)$$

should become independent random variables with densities $p_i(\varepsilon_{r,i})$ (7). Note that for any $\alpha > 0$ the limiting distribution is indeed given by (7), whereas (30) only holds for $\alpha = 0$. In contrast to the NPP for the MSP with $q > 2$ the bound $\alpha$ has to be strictly larger than zero because $g'(0) = 0$ for $q > 2$.

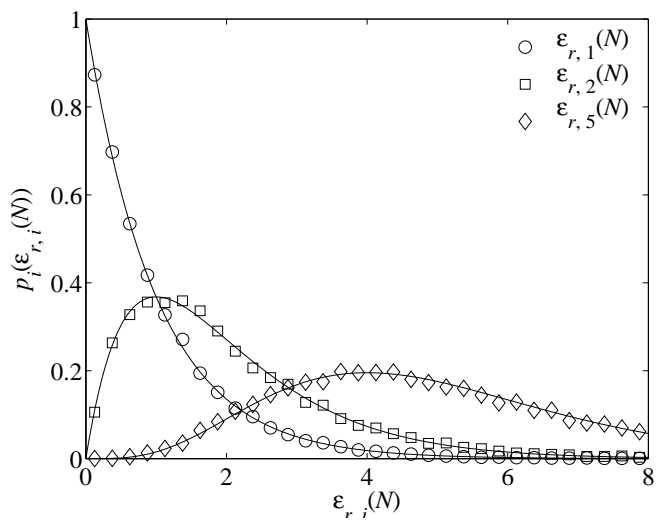

Figure 8: Distribution of the scaled energies for $\varepsilon_{r,i}(N)$ for $\alpha = 0.433$ and $q = 3$. Numerical simulations for $N = 20$ (symbols) and the predictions of the local random energy conjecture, (7) (lines). Each data point represents an average over 25 000 random instances.

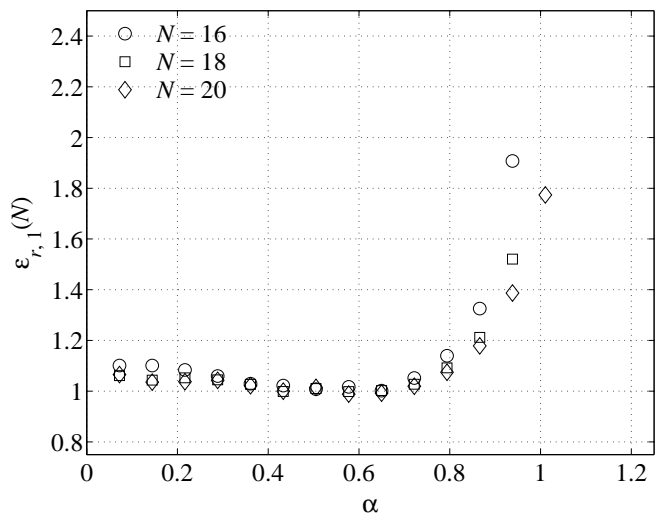

Figure 9: Mean $\overline{\varepsilon_{r,1}(N)}$ as a function of the parameter $\alpha$ for different system sizes $N$ for MSP with $q = 3$.

Figure 8 shows the numerical distribution of $\varepsilon_{r,i}(N)$ for $q = 3$ compared to the predictions of the local random energy conjecture. As in the case $q = 2$ the agreement is very good, and the strong finite size effects observed for $\alpha = 0$ are gone. If $\alpha$ gets too large (for finite $N$), more and more atypical configurations are sampled and deviations from the REM appear, see figure 9. This scenario too is very similar to the NPP with $q = 2$.

## 7. Conclusions and outlook

The first result of this paper is to reveal the local random energy scenario that is present in the energy spectrum of random number partitioning. We have given numerical evidence that adjacent energy levels and their corresponding configurations are uncorrelated. For the ground state energy and the first excitations this has been proven rigorously by Borgs, Chayes and Pittel [6]. It should be feasible to extent their proof to all but the largest energies, as formulated in our local REM conjecture in section 4. Generalising the BCP-Theorem to multiprocessor scheduling should also be doable. A mathematical analysis seems to be the only way to answer some of the questions our numerical approach has left open, like the question of the precise size of the neighbourhood in which energy levels are uncorrelated.

The second result is the missing correlation between energy and configurations that can be deduced from figure 5. As soon as one reaches energies smaller than $2/\sqrt{N}$, the cost function is essentially random, giving you no clue how to proceed to reach energies on the ground state scale $\mathcal{O}\left(2^{-N}\right)$. It is this REM nature of the NPP that accounts for the poor performance of heuristic algorithms [30, 31].

## Acknowledgments

This work was in part supported by the European Community's Human Potential program under contract No. HPRN-CT-2002-00319 STIPCO. H. B. and S. M. were supported by the German science council DFG under grant ME2044/1-1. Numerical simulations have been done on the Beowulf cluster TINA, see http://tina.nat.uni-magdeburg.de.